\documentclass[a4paper,preprint]{revtex4}
\usepackage{epsf}
\usepackage[dvips]{graphicx}
\usepackage{graphics}

\begin{document}
\title{What is the connection between ballistic deposition and the Kardar-Parisi-Zhang equation?}
\author{Eytan Katzav}
\email{eytak@post.tau.ac.il}
\author{Moshe Schwartz}
\email{mosh@tarazan.tau.ac.il}
\affiliation {School of Physics and Astronomy, Raymond and Beverly Sackler
Faculty of Exact Sciences, Tel Aviv University, Tel Aviv 69978, Israel}

\begin{abstract}
Ballistic deposition (BD) is considered to be a paradigmatic discrete growth model that represents the
Kardar-Parisi-Zhang (KPZ) universality class. In this paper we question this connection by rigorously deriving a
formal continuum equation from the BD microscopic rules, which deviates from the KPZ equation. In one dimension
these deviations are not important in the presence of noise, but for higher dimensions or when considering
deterministic evolution they are very relevant.
\end{abstract}

\maketitle

\section{Introduction}
Kinetic roughening of nonequilibrium surface growth has been of great interest in recent years. The kinetic
growth processes have been intensively studied via various discrete models and continuum equations and exhibit
nontrivial scaling behavior \cite{EW,barabasi95,meakin93,halpin95,krug97}. The surface width $W$, which is the
standard deviation of the surface height, scales as $W\left( {L,t} \right)\sim L^\alpha  g\left( {{t
\mathord{\left/ {\vphantom {t {L^z }}} \right. \kern-\nulldelimiterspace} {L^z }}} \right)$, where the scaling
function $g\left( u \right)$ is constant for $u \gg 1$ and behaves like $u^\beta$ for $u \ll 1$. Since the
growth is independent of the system size $L$ at the beginning of the process, the exponents must obey the
scaling relation $\beta  = {\alpha  \mathord{\left/ {\vphantom {\alpha  z}} \right. \kern-\nulldelimiterspace}
z}$. The scaling behavior of the growth is characterized by the roughness exponent $\alpha$, the growth exponent
$\beta$, and the dynamical exponent $z$, and these exponents determine the universality class.

Given a discrete model, it can be very difficult to assign to it a continuum equation that describes its
dynamics, and vice versa. Nowadays, this kind of problems is usually tackled using extensive simulations that
determine the critical exponents, by which the universality class is determined. A continuum system
characterized by the same exponents is then said to belong to the same universality class.

\begin{figure}[htb]
\includegraphics[width=8cm]{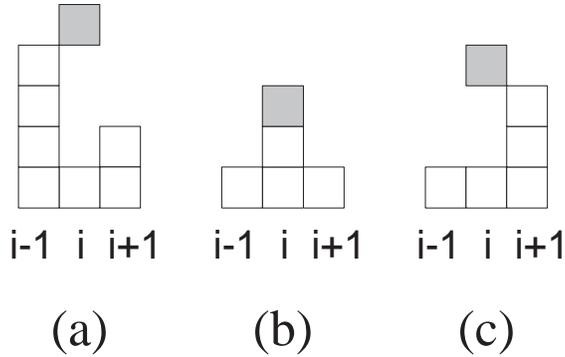}
\caption{Schematic representation of the next-nearest-neighbor ballistic deposition (NNN BD) model. A particle
falls vertically and sticks to the first site along its trajectory that has an occupied nearest neighbor. The
particle is allowed to stick to a diagonal neighbor as well.}
\label{BDrules}
\end{figure}

For example, the ballistic deposition (BD) model \cite{barabasi95}, which is a discrete model, is believed to be
described by the famous Kardar-Parisi-Zhang (KPZ) equation \cite{kpz86}. To be more specific, we will briefly
present these two models, since they will interest us in this paper. The ballistic deposition (BD) model in $1 +
1$ dimension (actually on a two-dimensional square lattice) can be described as follows. At time $t$, the height
of the interface of site $i$ is $h_i \left( t \right)$. We choose a random position above the surface and allow
a particle to fall vertically toward it. The particle sticks to the first site along the trajectory that has an
occupied nearest neighbor. If no such neighbor exists, it lands on the surface below. Actually, the version just
described is called NN (nearest neighbor) BD. In another version of the model, also known as NNN (next nearest
neighbor) BD \cite{barabasi95}, which will interest us in this paper, the particle is allowed to stick to a
diagonal neighbor as well, as shown in Fig. (\ref{BDrules}). At time $t + 1$, a column $i$ is chosen at random,
and the height $h_i \left( {t + 1} \right)$ is then given by
\begin{equation}
h_i \left( {t + 1} \right) = \max \left\{ {h_{i - 1} \left( t
\right),h_i \left( t \right),h_{i + 1} \left( t \right)}
\right\}+1
\label{1}.
\end{equation}
Not surprisingly, it turns out that both NN and NNN models are
well describes by the critical exponents of the KPZ equation, as
they share the same basic growth mechanism. Actually, even more
sophisticated versions of the model, including deposition of $n$
particles at a time, and a similar model that describes formation
of foam (see ref. \cite{rivier01}) show the same universal
behavior.

As mentioned above, the continuum equation that is believed to capture the essential dynamics of the BD models
is the famous KPZ \cite{kpz86} equation, given by
\begin{equation}
\frac{{\partial h}}{{\partial t}}\left( {\vec r,t} \right) = \nu \nabla ^2 h\left( {\vec r,t} \right) +
\frac{\lambda }{2}\left( {\nabla h} \right)^2  + \eta \left( {\vec r,t} \right)
\label{2},
\end{equation}
where $h\left( {\vec r,t} \right)$ is height of the interface above $\vec r$ at time $t$, and $\eta \left( {\vec
r,t} \right)$ is a noise term such that
\begin{eqnarray}
 \left\langle {\eta \left( {\vec r,t} \right)} \right\rangle  &=& 0 \nonumber\\
 \left\langle {\eta \left( {\vec r,t} \right)\eta \left( {\vec r',t} \right)} \right\rangle  &=& 2D_0 \delta \left( {\vec r - \vec r'} \right)\delta \left( {t - t'} \right)
\label{3}.
\end{eqnarray}

Many proposals for establishing more direct associations between discrete models and continuum equations of
motion are now present. First, there are phenomenological \cite{villain91} and symmetry
\cite{barabasi95,kpz86,hwa92,marsili96} arguments that can be very illuminating, but have limited power since
they cannot connect microscopic quantities to macroscopic ones. Sometimes it is possible to map the growth
problem onto other models \cite{kpz86, park01}, for which reliable results are known. A different approach,
based on real-space renormalization-group methods \cite{lam93}, is designed to identify the relevant macroscopic
constants (for example, diffusion coefficient) from the numerical data, once the universality class of the model
is already known.

However, in the last few years most of the efforts in establishing direct connections between discrete models
and continuum equations have been directed towards developing a general procedure via formal expansions of
discrete equations of motion
\cite{park01,racz91,vvedensky93,predota96,costanza97,park95,bantay92,corral97,tokihiro96,nagatani98,vvedensky03}.
Usually, the derivation of the continuum equation is based on regularizing and coarse graining discrete Langevin
equations that are obtained from a Kramers-Moyal expansion of the master equation. In simple words, transition
probabilities are calculated from the microscopic rules of the model for any given discrete height configuration
$\left\{ {h_i } \right\}$. These expressions usually contain discrete $\theta$ (Heaviside) and $\delta$ (Dirac)
functions. But, since the transition probabilities are supposed to be continuous functions (so that the
expansion that is used there is meaningful) some coarse-graining procedure is needed. More specifically, this
involves expansions of the form
\begin{equation}
\theta \left( x \right) = 1 + \sum\limits_{k = 1}^\infty  {A_k x^k }
\label{4},
\end{equation}
as originally suggested in \cite{vvedensky93}. Sometimes a less restricted form $\theta \left( x \right) =
\sum\limits_{k = 0}^\infty  {A_k x^k } $ is used. Other suggestions include
\begin{equation}
\theta \left( x \right) = {{\left[ {1 + \tanh \left( {Cx} \right)} \right]} \mathord{\left/ {\vphantom {{\left[
{1 + \tanh \left( {Cx} \right)} \right]} 2}} \right. \kern-\nulldelimiterspace} 2}
\label{5},
\end{equation}
where $C$ is an arbitrary positive parameter, with the exact
$\theta \left( x \right)$ function being obtained in the limit $C
\to \infty $ \cite{park95}. $C$ is then used in the expansion as
an uncontrolled parameter. Others \cite{predota96} use the shifted
form $\theta \left( x \right) = \mathop {\lim }\limits_{C \to
\infty } {{\left[ {1 + \tanh \left( {C\left\{ {x + \alpha }
\right\}} \right)} \right]} \mathord{\left/ {\vphantom {{\left[ {1
+ \tanh \left( {C\left\{ {x + \alpha } \right\}} \right)} \right]}
2}} \right. \kern-\nulldelimiterspace} 2}$ with $\alpha  \in
\left( {0,{\textstyle{1 \over 2}}} \right]$, or a modified version
using $\arctan \left( {Cx} \right)$ \cite{bantay92} or $erf\left(
{Cx} \right)$ \cite{tokihiro96} instead of $\tanh \left( {Cx}
\right)$ (each version has its advantages) in eq. (\ref{5}).

In some cases the master equation approach is problematic, like in
the case of deriving the Kardar-Parisi-Zhang (KPZ) equation
\cite{kpz86} from the discrete model known as Ballistic Deposition
(BD) \cite{costanza97}. Thus a similar approach, yet more
appropriate for that specific case was developed. The method is
based on dealing directly with the discrete Langevin equation
rather than with its associated master equation. Still, expansions
like
\begin{equation}
\theta \left( {x - a} \right) = \theta \left( x \right) + \sum\limits_{n = 1}^\infty  {\frac{{a^n }}{{n!}}\left.
{\frac{{\partial ^n \theta \left( y \right)}}{{\partial y^n }}} \right|_{y = x} }
\label{6}
\end{equation}
were used. A different, yet closely related Langevin based, approach used the following representation of the
$\max $ function \cite{tokihiro96,nagatani98}
\begin{equation}
\max \left\{ {A,B,C} \right\} = \mathop {\lim }\limits_{\varepsilon  \to 0^ +  } \varepsilon \ln \left\{ {e^{{A
\mathord{\left/
 {\vphantom {A \varepsilon }} \right.
 \kern-\nulldelimiterspace} \varepsilon }}  + e^{{B \mathord{\left/
 {\vphantom {B \varepsilon }} \right.
 \kern-\nulldelimiterspace} \varepsilon }}  + e^{{C \mathord{\left/
 {\vphantom {C \varepsilon }} \right.
 \kern-\nulldelimiterspace} \varepsilon }} } \right\}
\label{7},
\end{equation}
in order to go from the discrete BD model to the continuum KPZ equation. Lately \cite{vvedensky03}, influenced
by the last representation, the Edwards-Wilkinson \cite{EW} equation was derived from a discrete model using
\begin{equation}
\theta \left( x \right) = \max \left\{ {x + a,0} \right\} - \max \left\{ x \right\} = \mathop {\lim
}\limits_{\varepsilon  \to 0^ +  } \left\{ {\frac{\varepsilon }{a}\ln \left[ {\frac{{e^{{{\left( {x + a}
\right)} \mathord{\left/ {\vphantom {{\left( {x + a} \right)} \varepsilon }} \right. \kern-\nulldelimiterspace}
\varepsilon }}  + 1}}{{e^{{x \mathord{\left/ {\vphantom {x \varepsilon }} \right. \kern-\nulldelimiterspace}
\varepsilon }}  + 1}}} \right]} \right\}
\label{8},
\end{equation}
where a is any constant in the interval $\left( {0,1} \right]$.

In spite of these many new and interesting derivations, one can easily point out three main drawbacks of this
last approach. First, in many cases the derivation is performed in one dimension, where higher dimensions are
not discussed at all, or specifically known to cause fatal difficulties (see ref. \cite{nagatani98} for
example).

Second, obviously the mathematics used in the derivation is not so
"kosher". For example, an expansion like the one given in eq.
(\ref{4}) is problematic because the Heaviside function is
certainly not analytic around zero. Another example is when taking
an expression like eq. (\ref{5}) and expanding it for small $C$ -
while the limiting procedure that is needed for the equality to
hold requires $C \to \infty $.

Third, since artificial parameters like $C$ and $\epsilon$ that cannot always be removed later enter the
discussion it is not possible to infer the macroscopic quantities (such as the diffusion coefficient) from the
microscopic rules.

In this paper, we focus on the ballistic deposition (BD) model, and its believed relation with the
Kardar-Parisi-Zhang (KPZ) equation, since it is more difficult to establish \cite{costanza97,nagatani98} than
the relation of KPZ to other discrete models such as solid-on-solid (SOS). In what follows we will question this
relation, and claim that it is not a coincidence that such a formal (and "kosher") derivation was not found. In
section II we will show that strictly speaking the continuum model that describes BD in one dimension is not the
KPZ equation, but rather an equation with $\left| {\nabla h} \right|$ instead of the quadratic $\left( {\nabla
h} \right)^2 $ term. Then, we will claim that in the case of stochastically driven growth, the difference
between the two equations is not dramatic, and just modifies the value of the coupling constant in the KPZ
equation. In section III we show (using \cite{krug88,amar93}) that when the noise is shut down and the dynamics
becomes deterministic the difference between the BD and KPZ is very important. In section IV, we make a step
forward towards a derivation of a continuum equation that emanates from BD in higher dimensions. However, the
equation we find reflects the underlying structure of the discrete lattice, in the sense that the equation
depends on the directions of the coordinate system induced by the lattice. Therefore, the equation we obtain
cannot be conclusively related to the KPZ equation. At the end, in section V, a brief summary of the results
obtained in this paper is presented.

\section{Derivation of a continuum equation in one-dimension}

We begin with the following one-dimensional discrete model
\begin{equation}
h_i \left( {t + 1} \right) = \max \left\{ h_{i - 1} \left( t
\right),h_i \left( t \right),h_{i + 1} \left( t \right) \right\}+
\eta _i \left( t \right)
\label{9},
\end{equation}
where $h_i \left( t \right)$ is the height of a surface at lattice
position $i$ and time $t$ (the time is discrete as well). In
addition, $\eta _i \left( t \right)$ is a white noise term
satisfying
\begin{eqnarray}
 \left\langle {\eta _i \left( t \right)} \right\rangle  &=& 0 \nonumber\\
 \left\langle {\eta _i \left( t \right)\eta _{i'} \left( {t'} \right)} \right\rangle &=& 2D_0 \delta _{i,i'} \delta _{t,t'}
\label{10}.
\end{eqnarray}
This discrete model is a formal expression for an $n$-particle NNN (next nearest neighbor) BD model (see eq.
(21) in ref. \cite{nagatani98}). It is well known that the fact that here we deposit more than one particle per
unit time is not important for the universal behavior of the model. Actually the same discrete model is used to
describe formation of foam \cite{rivier01}.

First we represent the $\max \left\{ {} \right\}$ operator using Heaviside functions
\begin{equation}
\max \left\{ {a,b} \right\} = a\theta \left( {a - b} \right) + b\theta \left( {b - a} \right)
\label{11}.
\end{equation}
Thus, for three arguments we also get
\begin{equation}
\max \left\{ {a,b,c} \right\} = a\theta \left( {a - b} \right)\theta \left( {a - c} \right) + b\theta \left( {b
- a} \right)\theta \left( {b - c} \right) + c\theta \left( {c - a} \right)\theta \left( {c - b} \right)
\label{12}.
\end{equation}
Now we represent the Heaviside function using the sign function
\begin{equation}
\theta \left( x \right) = \frac{1}{2} + \frac{1}{2}{\mathop{\rm sgn}} \left( x \right)
\label{13}.
\end{equation}
Applying these representations to eq. (\ref{9}) we get
\begin{eqnarray}
&& \max \left\{ {h_{i - 1} \left( t \right),h_i \left( t \right),h_{i + 1} \left( t \right)} \right\} \nonumber\\
 &=& h_{i - 1} \left( t \right){\textstyle{1 \over 4}}\left[ {1 + {\mathop{\rm sgn}} \left( {h_{i - 1} \left( t \right) - h_i \left( t \right)} \right)} \right]\left[ {1 + {\mathop{\rm sgn}} \left( {h_{i - 1} \left( t \right) - h_{i + 1} \left( t \right)} \right)} \right] \nonumber \\
 &+& h_i \left( t \right){\textstyle{1 \over 4}}\left[ {1 + {\mathop{\rm sgn}} \left( {h_i \left( t \right) - h_{i - 1} \left( t \right)} \right)} \right]\left[ {1 + {\mathop{\rm sgn}} \left( {h_i \left( t \right) - h_{i + 1} \left( t \right)} \right)} \right] \nonumber \\
 &+& h_{i + 1} \left( t \right){\textstyle{1 \over 4}}\left[ {1 + {\mathop{\rm sgn}} \left( {h_{i + 1} \left( t \right) - h_{i - 1} \left( t \right)} \right)} \right]\left[ {1 + {\mathop{\rm sgn}} \left( {h_{i + 1} \left( t \right) - h_i \left( t \right)} \right)} \right]
\label{14}.
\end{eqnarray}
Thus, after some simple algebra
\begin{eqnarray}
 h_i \left( {t + 1} \right) - h_i \left( t \right) &=& {\textstyle{1 \over 4}}\left[ {h_{i - 1} \left( t \right) - 2h_i \left( t \right) + h_{i + 1} \left( t \right)} \right] - {\textstyle{1 \over 4}}h_i \left( t \right) \nonumber\\
 \quad \quad  &+& {\textstyle{1 \over 4}}\left[ {h_i \left( t \right) - h_{i - 1} \left( t \right)} \right]{\mathop{\rm sgn}} \left( {h_i \left( t \right) - h_{i - 1} \left( t \right)} \right) \nonumber \\
 \quad \quad  &+& {\textstyle{1 \over 4}}\left[ {h_{i + 1} \left( t \right) - h_{i - 1} \left( t \right)} \right]{\mathop{\rm sgn}} \left( {h_{i + 1} \left( t \right) - h_{i - 1} \left( t \right)} \right) \nonumber \\
 \quad \quad  &+& {\textstyle{1 \over 4}}\left[ {h_{i + 1} \left( t \right) - h_i \left( t \right)} \right]{\mathop{\rm sgn}} \left( {h_{i + 1} \left( t \right) - h_i \left( t \right)} \right) \nonumber \\
 \quad \quad  &+& {\textstyle{1 \over 4}}h_{i - 1} \left( t \right){\mathop{\rm sgn}} \left( {h_i \left( t \right) - h_{i - 1} \left( t \right)} \right){\mathop{\rm sgn}} \left( {h_{i + 1} \left( t \right) - h_{i - 1} \left( t \right)} \right) \nonumber \\
 \quad \quad  &-& {\textstyle{1 \over 4}}h_i \left( t \right){\mathop{\rm sgn}} \left( {h_i \left( t \right) - h_{i - 1} \left( t \right)} \right){\mathop{\rm sgn}} \left( {h_{i + 1} \left( t \right) - h_i \left( t \right)} \right) \nonumber \\
 \quad \quad  &+& {\textstyle{1 \over 4}}h_{i + 1} \left( t \right){\mathop{\rm sgn}} \left( {h_{i + 1} \left( t \right) - h_{i - 1} \left( t \right)} \right){\mathop{\rm sgn}} \left( {h_{i + 1} \left( t \right) - h_i \left( t \right)} \right) + \eta _i \left( t \right)
\label{15}.
\end{eqnarray}
The last expression is written in such a way that identifying
discrete derivatives is easy. Therefore, denoting the spatial and
temporal increments by $\Delta x$ and $\Delta t$, and using the
simplifying fact ${\mathop{\rm sgn}} \left( {ax} \right) =
{\mathop{\rm sgn}} \left( x \right)$ (when $a > 0$) we get
\begin{equation}
\frac{{\partial h}}{{\partial t}}\left( {x,t} \right) = \frac{{\left( {\Delta x} \right)^2 }}{{4\Delta t}}\left[
{1 + {\mathop{\rm sgn}} ^2 \left( {\frac{{\partial h}}{{\partial x}}} \right)} \right]\frac{{\partial ^2
h}}{{\partial x^2 }}\left( {x,t} \right) + \frac{{\Delta x}}{{\Delta t}}\left| {\frac{{\partial h}}{{\partial
x}}} \right| + \frac{1}{{\Delta t}}\eta \left( {x,t} \right)
\label{16}.
\end{equation}
Note that ${\rm sgn}^2 \left( x \right) = \left\{ \begin{array}{l} 1\quad \quad x \ne 0 \\ 0\quad \quad x = 0
\end{array} \right.$. Thus, for a strictly non smooth surface (i.e. a surface that is not flat), almost
everywhere we can use the replacement ${\mathop{\rm sgn}} ^2 \left( {{\textstyle{{\partial h} \over {\partial
x}}}} \right) = 1$ in order to further simplify eq. (\ref{16}). In addition, since in the discrete model we
actually took $\Delta x = \Delta t = 1$ we get the final result
\begin{equation}
\frac{{\partial h}}{{\partial t}}\left( {x,t} \right) = \frac{1}{2}\frac{{\partial ^2 h}}{{\partial x^2 }}\left(
{x,t} \right) + \left| {\frac{{\partial h}}{{\partial x}}} \right| + \eta \left( {x,t} \right)
\label{17}.
\end{equation}

This equation is a result of a straightforward exact
coarse-graining of eq. (\ref{9}). Thus, one can be convinced that
BD (at least this $n$-particle version of it) does not lead
exactly to the KPZ equation. This fact might explain the
difficulties encountered in the past while deriving the KPZ
equation from BD.

At this point, it is interesting to wonder what is the connection between BD and the KPZ equation? After all,
for the last two decades BD is considered to be a paradigmatic discrete growth model that belongs to the KPZ
universality class (see for example the approach taken by Barab\'asi and Stanley \cite{barabasi95} to present
the field of surface growth). In addition, extensive simulations
\cite{rivier01,Family85,meakin86,baiod88,reis01,ko94,family90} aimed at extracting the KPZ critical exponents
used BD, and found critical exponents that are consistent with other known results. However, one important point
should be emphasized. Most of the simulations \cite{Family85,meakin86,baiod88,reis01,ko94,family90} that used BD
were done in one and two dimensions, while only one of them (ref. \cite{ko94}) studied BD in three dimensions.
This point is important because in one and two dimensions only a rough phase is possible, while for higher
dimensions a phase transition between a smooth phase and a rough phase becomes possible. The last remark will be
a starting point for the discussion in section IV, where the behavior of BD in higher dimensions in discussed.
In the following, we will discuss, however, the equivalence of the one dimensional continuous model (eq.
\ref{17}) we derived from the BD deposition discrete model. A support to that equivalence is already found in
ref. \cite{amar93} that studied stochastically driven Langevin equations (of the KPZ type) with a generalized
$\left| {\nabla h} \right|^\mu$ nonlinearity. They concluded that no-matter what value of $\mu$ was taken, KPZ
behavior was revealed, as long as the noise term is present (the case of no noise is interesting by itself, and
will be discussed in section III below). This result should not be a surprise nowadays as it is consistent with
the well known symmetry argument \cite{barabasi95,kpz86,hwa92,marsili96} that is often used to justify the KPZ
equation -namely since the generalized nonlinearity $\left| {\nabla h} \right|^\mu  $ breaks the $h \to - h$
symmetry it should belong to the KPZ universality class. Thus, this fundamental observation of Amar and Family
\cite{amar93} establishes the link between eq. (\ref{17}) and the KPZ equation, by simply identifying $\mu=1$ in
that equation. Hence, the route between BD and KPZ in one dimension is now understood, and the common knowledge
about BD is justified.

Now, it would be interesting to complete the picture and to extract the macroscopic coefficients that describe
the continuum KPZ equation that is related to the BD model. First, the diffusion coefficient can be inferred
easily from equation (\ref{17}). Then, the noise amplitude is unaffected by the coarse graining. Finally, we
would like to identify the coupling constant $\lambda $ in the KPZ equation (\ref{2}). However, since the
nonlinear term in eq. (\ref{17}) is not in a KPZ form, $\lambda $ cannot be just read from the equation. Thus,
we push further the idea presented in the previous paragraph, and make the following identification
\begin{equation}
\frac{{\left| {\nabla h} \right|}}{{\left\langle {\left| {\nabla h} \right|} \right\rangle }} \simeq
\frac{{\left| {\nabla h} \right|^2 }}{{\left\langle {\left| {\nabla h} \right|^2 } \right\rangle }}
\label{18},
\end{equation}
where $\left\langle  \cdots  \right\rangle $ means steady-state averaging. The last equation leads to$\left|
{\nabla h} \right| \simeq {\textstyle{{\left\langle {\left| {\nabla h} \right|} \right\rangle } \over
{\left\langle {\left| {\nabla h} \right|^2 } \right\rangle }}}\left( {\nabla h} \right)^2 $. Plugging this
estimate into eq. (\ref{17}), gives the prediction $\lambda/2 \simeq {\textstyle{{\left\langle {\left| {\nabla
h} \right|} \right\rangle } \over {\left\langle {\left| {\nabla h} \right|^2 } \right\rangle }}}$ for the
coupling constant of the KPZ equation (\ref{2}). The meaning of this expression is that the KPZ coupling
constant that describes the surface that grows under BD is obtained by calculating the two quantities
$\left\langle {\left| {\nabla h} \right|} \right\rangle $ and $\left\langle {\left| {\nabla h} \right|^2 }
\right\rangle $ in a BD simulation (in steady state) and plugging the obtained averages into the expression
given above for $\lambda/2$.

The last prediction has been checked numerically by us. We grew a one dimensional interface on a discrete
lattice with linear size $L = 1024$ using the microscopic rules of BD (given by eq. (\ref{9}) above). Using this
simulation we found $\left\langle {\left| {\nabla h} \right|} \right\rangle  = 0.4705$ and $\left\langle {\left|
{\nabla h} \right|^2 } \right\rangle  = 0.4639$ that leads to the prediction $\lambda  \simeq 2.17$. Using an
inverse method - that is "guessing" the best macroscopic parameters that recover the same growth process (see
refs. \cite{barabasi95,lam93}), we found $\lambda  = 2.27$, that supports the rough estimate of $\lambda $ we
suggested above.

Up to here we focused on the common characteristics of BD and KPZ. However, as will be seen in the following two
sections, if the noise is turned off, or if we go to higher dimensions, significant differences between the two
models can be found.

\section{Deterministic flattening}

In this section we discuss the dynamics of the discrete BD model and the continuum model with $\left| {\nabla h}
\right|^\mu$ non-linearity for $\mu=2$ (KPZ) and for $\mu=1$, when the noise term that appears in them is
eliminated. This limit mimics the physical scenario when the deposited materials stop falling, and the interface
relaxes to a flat surface, due to its diffusive term. We think that it is worth studying such physical
situations because it emphasizes the difference between the dynamics of BD and KPZ already in the
one-dimensional case.

\begin{figure}[htb]
\includegraphics[width=7cm]{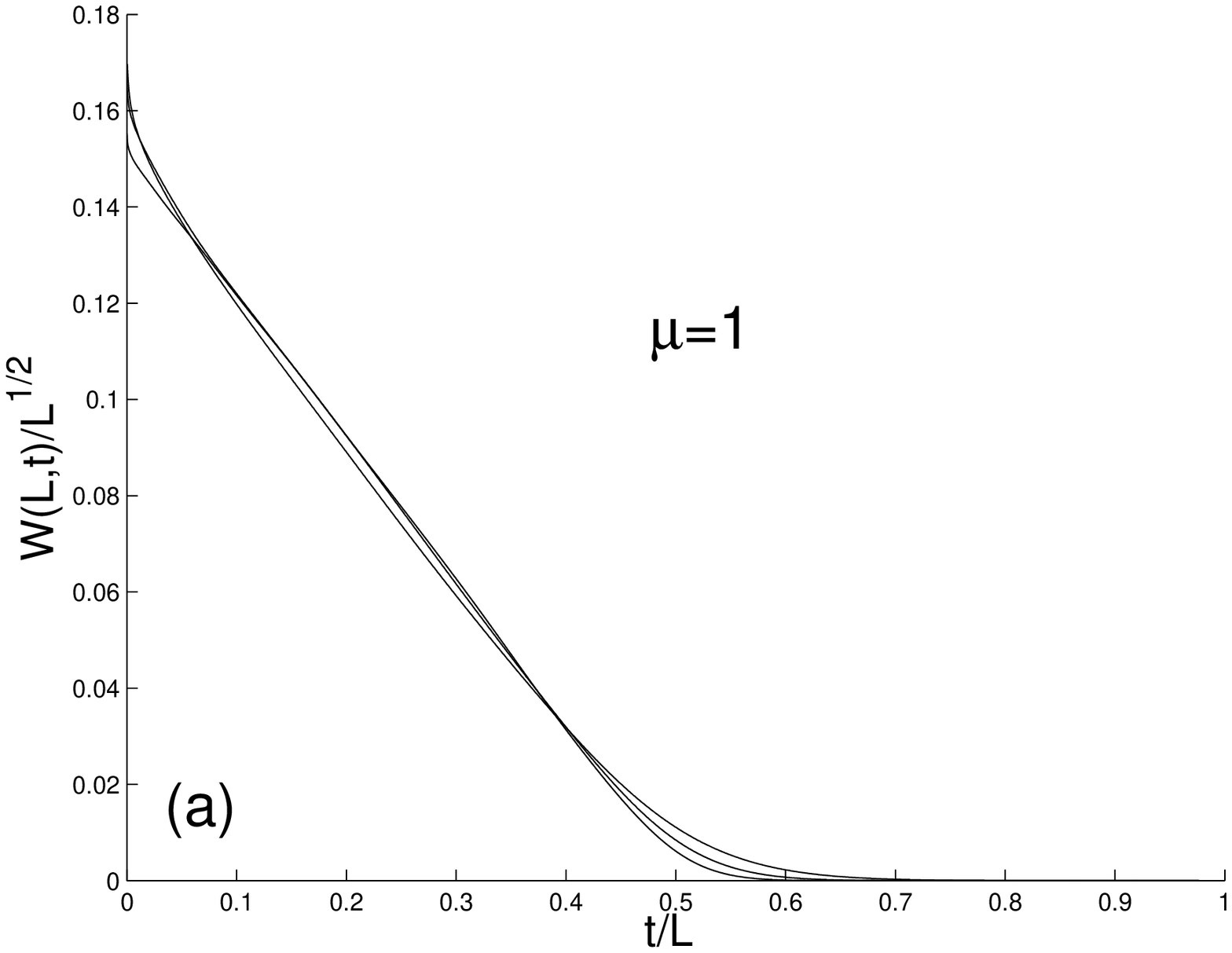}
\includegraphics[width=7cm]{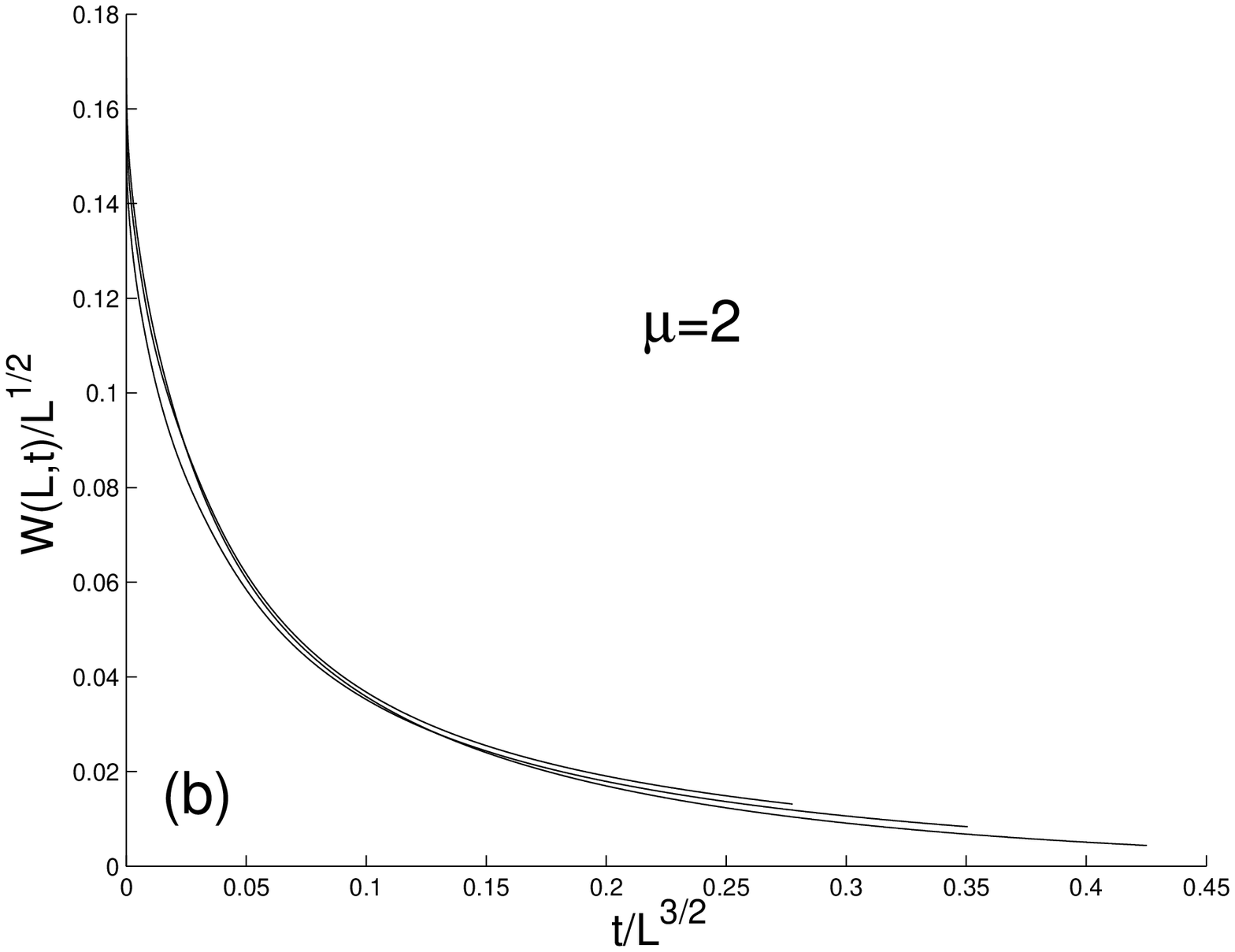}
\caption{Scaling plot for the deterministic equation (\ref{19}) for $L = 256,512$ and $1024$ with (a) $\mu=1$
using $z_d=1$ and with (b) $\mu  = 2$ using $z_d = {\textstyle{3 \over 2}}$.} \label{det1}
\end{figure}

The starting point of this discussion is the observation of Krug and Spohn \cite{krug88} regarding a
deterministic continuum equation of the general form
\begin{equation}
\frac{{\partial h}}{{\partial t}} = \nu \nabla ^2 h + g \left| {\nabla h} \right|^\mu \label{19},
\end{equation}
(where $\mu  \ge 1$) that describes the smoothing of an initially rough surface under deterministic growth. They
argued that if the initial surface has roughness exponent $\alpha$, the scaling relation
\begin{equation}
z_d = \min \left\{ {2,\mu \left( {1 - \alpha } \right) + \alpha } \right\} \label{20}
\end{equation}
should hold, where $z_d$ is a dynamic exponent for deterministic evolution, which similar to the stochastically
driven case, determines the time scale of the decaying width (see Fig. \ref{det1}).

Influenced by this work, Amar and Family \cite{amar93} made a systematic study of such surface growth equations
with a generalized nonlinearity and found good agreement between the scaling relation and the numerical
integration of equation (\ref{19}). We also performed a numerical integration, using the same integration scheme
as in ref. \cite{amar93} (the coefficients $\nu$ and $g$ were taken from eq. (\ref{17}) and using the result of
the previous section) for the cases $\mu  = 1$ (eq. (\ref{17})) and $\mu  = 2$ (KPZ) in 1+1 dimension. We also
used a rough surface with $\alpha  = {\textstyle{1 \over 2}}$ as an initial condition. A scaling plot of our
results for the surface width $W\left( {L,t} \right)$ for $\mu  = 1$ versus ${t \mathord{\left/ {\vphantom {t
{L^z_d}}} \right. \kern-\nulldelimiterspace} {L^z_d }}$ with $z_d = 1$ is given in Fig. \ref{det1}(a), and a
similar scaling plot for $\mu  = 2$ with $z_d = {\textstyle{3 \over 2}}$ is shown in Fig. \ref{det1}(b).

As one can see, the scaling is good in both cases, in agreement
with the scaling relation (\ref{20}) and therefore consistent with
refs. \cite{krug88,amar93}. Thus, the difference between the
deterministic dynamics of the KPZ equation and that of eq.
(\ref{17}) is now self evident, and not new. Two questions should
be asked here. Our derivation of eq. (\ref{17}) does not depend on
the presence of noise. Therefore, we might expect that even when
deterministic dynamics is considered discrete BD will be
equivalent to the $\mu=1$ case rather than to the $\mu=2$ case
(KPZ). The first question is whether this is indeed the case. The
second question regards the predicted values for the diffusion
coefficient (namely $\nu  = {\textstyle{1 \over 2}}$) and the
coupling (i.e., the coefficient of the $\left| {\nabla h}
\right|^\mu$ term - namely $g=1$) for the $n$-particle NNN BD
model that we read from equation (\ref{17}). The question is
whether this prediction is a sensible one?

\begin{figure}[htb]
\includegraphics[width=7cm]{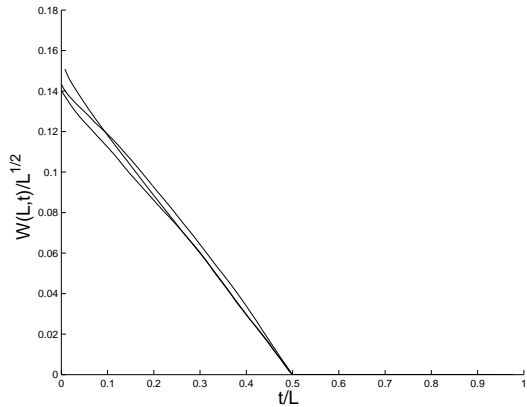}
\caption{Scaling plot for the deterministic NNN BD model given by eq. (\ref{9}) for $L=256,512$ and $1024$ using
$z_d = 1$.} \label{det2}
\end{figure}

In order to answer these questions we implemented the NNN BD model (whose sticking rules are given by eq.
(\ref{9})) on a two dimensional square lattice ($1 + 1$ dimensions), and measured the surface width $W\left(
{L,t} \right)$ of the smoothing surface, using a rough surface with $\alpha  = {\textstyle{1 \over 2}}$ as an
initial condition (just like the continuum models above). We present the results in Fig. \ref{det2}.

As can be seen in Fig. \ref{det2}, the deterministic dynamics of the discrete model is evidently the one we have
seen for the $\mu  = 1$ continuum equation, since the data collapse in the scaling plot with $z_d = 1$ is good.
Moreover, Fig. \ref{det1}(a) and Fig. \ref{det2} look pretty much the same. The main difference between the two
is the rounded tail in Fig. \ref{det1}(a) around ${t \mathord{\left/ {\vphantom {t L}} \right.
 \kern-\nulldelimiterspace} L} = 0.5$ compared to the sharp transition in Fig. \ref{det2}.
 This difference can be accounted for by the fact that the numerical integration of the continuum equation
 used a small time increment $\left( {\Delta t = 0.05} \right)$ while the discrete growth model used a much
 larger one $\left( {\Delta t = 1} \right)$. This implies a better temporal resolution around ${t \mathord{\left/
 {\vphantom {t L}} \right. \kern-\nulldelimiterspace} L} = 0.5$ by the numerical integration when compared
 to the simulation. I addition, the similarity between the two scaling plots means that the predicted macroscopic
 quantities are consistent with the results obtained from the
 simulation.

In this section we showed that a difference between BD model and
KPZ equation can be seen, when looking at their
 deterministic dynamics. We also showed that eq. (\ref{17}) captures the dynamical behavior of BD even in the
 deterministic regime.

\section{Results for higher dimensions}

In this section we generalize the formal derivation given in section II to higher dimensions. As will be seen,
the derivation is not exactly the same, and the resulting continuum equation is not a simple generalization of
the one dimensional equation (\ref{17}).

We begin with the  $n$-particle NNN BD model in $d$ dimensions (discretized on a cubic hyperlattice)
\begin{equation}
h_{\vec r} \left( {t + 1} \right) = \max \left\{ {h_{\vec r} \left( t \right),h_{\vec r + \hat x_1 } \left( t
\right),h_{\vec r - \hat x_1 } \left( t \right), \ldots ,h_{\vec r + \hat x_d } \left( t \right),h_{\vec r -
\hat x_d } \left( t \right)} \right\} + \eta _{\vec r} \left( t \right) \label{21},
\end{equation}
where $\hat x_i$ is a unit vector in the $i$'th direction.

Like in section II, using equations (\ref{12})-(\ref{13}), coarse graining (space) and using the simplifications
${\mathop{\rm sgn}} \left( {ax} \right) = {\mathop{\rm sgn}} \left( x \right)$ (for $a \ne 0$) and ${\mathop{\rm
sgn}} ^2 \left( {{\textstyle{{\partial h} \over {\partial x_i }}}} \right) = 1$ (which is again true almost
anywhere) we get
\begin{eqnarray}
&& 2^{2d} h_{\vec r} \left( {t + 1} \right) = 2\sum\limits_{i = 1}^d {\left\{ {h_{\vec r + \hat x_i } \left( t
\right)\left[ {1 + {\mathop{\rm sgn}} \left( {{\textstyle{{\partial h} \over {\partial x_i }}}} \right)}
\right]\prod\limits_{\scriptstyle j = 1 \hfill \atop
  \scriptstyle j \ne i \hfill}^d {\left[ {1 + {\mathop{\rm sgn}} \left( {{\textstyle{{\partial h} \over {\partial x_i }}} - {\textstyle{{\partial h} \over {\partial x_j }}}} \right)} \right]\left[ {1 + {\mathop{\rm sgn}} \left( {{\textstyle{{\partial h} \over {\partial x_i }}} + {\textstyle{{\partial h} \over {\partial x_j }}}} \right)} \right]}  + } \right.}  \nonumber\\
&&\left. { + h_{\vec r - \hat x_i } \left( t \right)\left[ {1 - {\mathop{\rm sgn}} \left( {{\textstyle{{\partial
h} \over {\partial x_i }}}} \right)} \right]\prod\limits_{\scriptstyle j = 1 \hfill \atop
  \scriptstyle j \ne i \hfill}^d {\left[ {1 - {\mathop{\rm sgn}} \left( {{\textstyle{{\partial h} \over {\partial x_i }}} - {\textstyle{{\partial h} \over {\partial x_j }}}} \right)} \right]\left[ {1 - {\mathop{\rm sgn}} \left( {{\textstyle{{\partial h} \over {\partial x_i }}} + {\textstyle{{\partial h} \over {\partial x_j }}}} \right)} \right]} } \right\} + \eta _{\vec r} \left( t \right)
\label{22}.
\end{eqnarray}
Using $1 \pm sgn\left( x \right) = 2\theta \left( { \pm x} \right)$ we get
\begin{eqnarray}
 h_{\vec r} \left( {t + 1} \right) &=& \sum\limits_{i = 1}^d {\left\{ {h_{\vec r + \hat x_i } \left( t \right)\theta \left( {{\textstyle{{\partial h} \over {\partial x_i }}}} \right)\prod\limits_{\scriptstyle j = 1 \hfill \atop
  \scriptstyle j \ne i \hfill}^d {\left[ {\theta \left( {{\textstyle{{\partial h} \over {\partial x_i }}} - {\textstyle{{\partial h} \over {\partial x_j }}}} \right)\theta \left( {{\textstyle{{\partial h} \over {\partial x_i }}} + {\textstyle{{\partial h} \over {\partial x_j }}}} \right)} \right]}  + } \right.}  \nonumber\\
&&\left. { + h_{\vec r - \hat x_i } \left( t \right)\theta \left( { - {\textstyle{{\partial h} \over {\partial
x_i }}}} \right)\prod\limits_{\scriptstyle j = 1 \hfill \atop
  \scriptstyle j \ne i \hfill}^d {\left[ {\theta \left( { - {\textstyle{{\partial h} \over {\partial x_i }}} + {\textstyle{{\partial h} \over {\partial x_j }}}} \right)\theta \left( { - {\textstyle{{\partial h} \over {\partial x_i }}} - {\textstyle{{\partial h} \over {\partial x_j }}}} \right)} \right]} } \right\} + \hat \eta _{\vec r} \left( t \right)
\label{23},
\end{eqnarray}
where $\hat \eta _{\vec r} \left( t \right) = 2^{ - 2d} \eta _{\vec r} \left( t \right)$. Notice that
\begin{equation}
\theta \left( a \right)\theta \left( {a + x} \right)\theta \left( {a - x} \right) = \theta \left( a
\right)\theta \left( {a - \left| x \right|} \right) = \theta \left( a \right)\theta \left( {\left| a \right| -
\left| x \right|} \right)
\label{24},
\end{equation}
thus
\begin{equation}
h_{\vec r} \left( {t + 1} \right) = \sum\limits_{i = 1}^d {\prod\limits_{\scriptstyle j = 1 \hfill \atop
  \scriptstyle j \ne i \hfill}^d {\theta \left( {\left| {{\textstyle{{\partial h} \over {\partial x_i }}}} \right| - \left| {{\textstyle{{\partial h} \over {\partial x_j }}}} \right|} \right)} \left[ {h_{\vec r + \hat x_i } \left( t \right)\theta \left( {{\textstyle{{\partial h} \over {\partial x_i }}}} \right) + h_{\vec r - \hat x_i } \left( t \right)\theta \left( { - {\textstyle{{\partial h} \over {\partial x_i }}}} \right)} \right]}  + \hat \eta _{\vec r} \left( t \right)
\label{25}.
\end{equation}
Using once again the relation $\theta \left( x \right) = {\textstyle{1 \over 2}}\left[ {1 + {\mathop{\rm sgn}}
\left( x \right)} \right]$, and reorganizing the last equation lead us to
\begin{eqnarray}
 h_{\vec r} \left( {t + 1} \right) &=& {\textstyle{1 \over 2}}\sum\limits_{i = 1}^d {\prod\limits_{\scriptstyle j = 1 \hfill \atop
  \scriptstyle j \ne i \hfill}^d {\theta \left( {\left| {{\textstyle{{\partial h} \over {\partial x_i }}}} \right| - \left| {{\textstyle{{\partial h} \over {\partial x_j }}}} \right|} \right)} }  \nonumber\\
 &\times &\left[ {h_{\vec r + \hat x_i } \left( t \right) - 2h_{\vec r} \left( t \right) + h_{\vec r - \hat x_i }
\left( t \right) + 2{\textstyle{{\partial h} \over {\partial x_i }}}{\mathop{\rm sgn}} \left(
{{\textstyle{{\partial h} \over {\partial x_i }}}} \right) + 2h_{\vec r} \left( t \right)} \right] + \hat \eta
_{\vec r} \left( t \right)
\label{26},
\end{eqnarray}
Coarse graining in both space and time gives us the final result
\begin{equation}
\frac{{\partial h}}{{\partial t}}\left( {\vec r,t} \right) = \sum\limits_{i = 1}^d {\left(
{\frac{1}{2}\frac{{\partial ^2 h}}{{\partial x_i^2 }} + \left| {\frac{{\partial h}}{{\partial x_i }}} \right|}
\right)\prod\limits_{\scriptstyle j = 1 \hfill \atop  \scriptstyle j \ne i \hfill}^d {\theta \left( {\left|
{{\textstyle{{\partial h} \over {\partial x_i }}}} \right| - \left| {{\textstyle{{\partial h} \over {\partial
x_j }}}} \right|} \right)} }  + \hat \eta \left( {\vec r,t} \right)
\label{27}.
\end{equation}

This equation might seem messy for a second. However, it has a
very simple structure. First, for $d = 1$ it collapses trivially
to eq. (\ref{17}). Then, in higher dimensions, the expression
$\prod\limits_{\scriptstyle j = 1 \hfill \atop   \scriptstyle j
\ne i \hfill}^d {\theta \left( {\left| {{\textstyle{{\partial h}
\over {\partial x_i }}}} \right| - \left| {{\textstyle{{\partial
h} \over {\partial x_j }}}} \right|} \right)}$ picks the
direction,$m$, in space along which the gradient is maximal, and
the considered part of the local growth rate has a contribution
from that direction alone, ${\textstyle{1 \over
2}}{\textstyle{{\partial ^2 h} \over {\partial x_m^2 }}} + \left|
{{\textstyle{{\partial h} \over {\partial x_m }}}} \right|$.
Actually, taking a glance at the original discrete model, this
result is not such a surprise since both formulations contain this
"maximal" growth ingredient in them.

Actually, the equation we obtain still reflects the directions
that are induced by the underlying lattice. Therefore, strictly
speaking, this is not a proper continuum description. Moreover, it
is not clear whether such a proper continuum description exists,
and how can it be derived. Therefore, the relation of this
equation to the $d$-dimensional KPZ equation for $d \ge 2$ is also
not evident. We leave these questions for future research.

\section{Summary and conclusions}

In this paper we discussed the connection between a discrete growth model, the next-nearest-neighbor ballistic
deposition, and a continuum equation, the Kardar Parisi-Zhang equation. It has been believed for the last two
decades \cite{barabasi95}, that BD is a paradigmatic discrete model for the KPZ universality class; however a
formal derivation was lacking. In this work, we show that the absence of a formal derivation is not accidental,
but rather reflects significant differences between the continuum equation that describes the BD model, and the
KPZ equation. This difference turns out to be mild in one-dimension in the presence of noise, but crucial when
discussing the deterministic dynamics. The relation between BD and KPZ is also questioned when going to higher
dimensions.

The main advantage of our approach is that it is direct and allows
to estimate of the macroscopic quantities of interest (such as the
diffusion coefficient) from the microscopic rules.

The main disadvantage of our approach is that it does not give a general prescription for deriving continuum
equations from discrete models. However, we do show that the option of giving a rigorous formal derivation
should not be overlooked. As we show in this paper, the usage of unjustified expansions (such as those presented
in eqs. (\ref{4})-(\ref{8}) above) might sometimes lead to wrong conclusions.

Two open questions are left for future work. First, whether the BD
model in $d$ dimensions has a proper continuum description, which
does not depend on the discrete lattice on which it is defined,
and how can one derive this equation. Sceond, whether the BD model
belongs to the KPZ universality class in dimensions higher than
one.

Acknowledgement: One of us (M.S.) would like to thank C. Oguey and
N. Rivier. The discussion with them on their model of foam growth
motivated our paper. M.S. would also like to thank the Isaac
Newton Institute where this discussion took place during their
program on foams and minimal surfaces.


\newpage


\begin{thebibliography}{41}
\bibitem{EW} S.F. Edwards and D. R. Wilkinson,
  {\it Proc. R. Soc. London Ser. A {\bf 381}}, 17 (1982).
\bibitem{barabasi95} A.-L. Barab\'asi and H. E. Stanley, Fractal Concepts in Surface Growth (Cambridge Univ. Press,
Cambridge, 1995).
\bibitem{meakin93} P. Meakin,
  {\it Phys. Rep. {\bf 235}}, 189 (1993).
\bibitem{halpin95} T. Halpin-Healy and Y.-C. Zhang,
  {\it Phys. Rep. {\bf 254}}, 215 (1995).
\bibitem{krug97} J. Krug,
  {\it Adv. in Phys. {\bf 46}}, 139 (1997).
\bibitem{rivier01} C. Oguey and N. Rivier,
  {\it J. Phys. A {\bf 34}}, 6225 (2001).
\bibitem{villain91}  J. Villain,
  {\it J. Phys. I {\bf 1}}, 19 (1991).
\bibitem{kpz86} M. Kardar, G. Parisi and Y.-C. Zhang,
  {\it Phys. Rev. Lett. {\bf 56}}, 889 (1986).
\bibitem{hwa92} T. Hwa and M. Kardar,
  {\it Phys. Rev. A {\bf 45}}, 7002 (1992).
\bibitem{marsili96} M. Marsili, A. Maritan, F. Toigo, and J. R. Banavar,
  {\it Rev. Mod. Phys. {\bf 68}}, 963 (1996).
\bibitem{park01} S.-C. Park, D. Kim, and J.-M. Park,
  {\it Phys. Rev. E {\bf 65}}, 015102(R) (2001).
\bibitem{lam93} C.-H. Lam and L.M. Sander,
  {\it Phys. Rev. Lett. {\bf 71}}, 561 (1993).
\bibitem{racz91} Z. R\'acz, M. Siegert, D. Liu, and M. Plischke,
  {\it Phys. Rev. A {\bf 43}}, 5275 (1991).
\bibitem{vvedensky93} D.D. Vvedensky, A. Zangwill, C.N. Luse, and M.R. Wilby,
  {\it Phys. Rev. E {\bf 48}}, 852 (1993).
\bibitem{predota96} M. Predota and M. Kotrla,
  {\it Phys. Rev. E {\bf 54}}, 3933 (1996).
\bibitem{costanza97} G. Costanza,
  {\it Phys. Rev. E {\bf 55}}, 6501 (1997).
\bibitem{park95} K. Park and B.N. Kahng,
  {\it Phys. Rev. E {\bf 51}}, 796 (1995).
\bibitem{bantay92} P. Bantay and I.M. Janosi,
  {\it Phys. Rev. Lett. {\bf 68}}, 2058 (1992).
\bibitem{corral97} A. Corral and A. Diaz-Guilera,
  {\it Phys. Rev. E {\bf 55}}, 2434 (1997).
\bibitem{tokihiro96} T. Tokihiro, D. Takahashi, J. Matsukidaira, and J. Satsuma,
  {\it Phys. Rev. Lett. {\bf 76}}, 3247 (1996).
\bibitem{nagatani98} T. Nagatani,
  {\it Phys. Rev. E {\bf 58}}, 700 (1998).
\bibitem{vvedensky03} D. D. Vvedensky,
  {\it Phys. Rev. E {\bf 67}}, 25102(R) (2003).
\bibitem{krug88} J. Krug and H. Spohn,
  {\it Phys. Rev. A {\bf 38}}, 4271 (1988).
\bibitem{amar93} J. G. Amar and F. Family,
  {\it Phys. Rev. E {\bf 47}}, 1595 (1993).
\bibitem{Family85} F. Family and T. Vicsek,
 {\it J. Phys. A {\bf 18}}, L75 (1985).
\bibitem{meakin86} P. Meakin, P. Ramanlal, L. M. Sander and R.C. Ball,
  {\it Phys. Rev. A {\bf 34}}, 5091 (1986).
\bibitem{baiod88} R. Baiod, D. Kessler, P. Ramanlal, L. Sander and R. Savit,
  {\it Phys. Rev. A {\bf 38}}, 3672 (1988).
\bibitem{reis01} F.D.A.A. Reis,
  {\it Phys. Rev. E {\bf 63}}, 056116 (2001).
\bibitem{ko94} D.Y.K. Ko and F. Seno,
  {\it Phys. Rev. E {\bf 50}}, 1741 (1994).
\bibitem{family90} F. Family,
 {\it Physica A {\bf 168}}, 561 (1990).
\bibitem{UCD98} T. Ala-Nissila,
  {\it Phys. Rev. Lett. {\bf 80}}, 887 (1998).
  \\ J. M. Kim, {\it ibid. {\bf 80}}, 888 (1998).
  \\ M. Lassig and H. Kinzelbach, {\it ibid. {\bf 80}}, 889 (1998).
\bibitem{Perlsman96} E. Perlsman and M. Schwartz,
 {\it Physica A {\bf 234}}, 523 (1996).
\bibitem{Castellano98} C. Castellano, M. Marsili and L. Pietronero,
  {\it Phys. Rev. Lett. {\bf 80}}, 3527 (1998).
  \\ C. Castellano, A. Gabrielli, M. Marsili, M.A. Munoz and L. Pietronero,
  {\it Phys. Rev. E {\bf 58}}, 5209 (1998).
\bibitem{Halpin90} T. Halpin-Healy,
  {\it Phys. Rev. A {\bf 42}}, 711 (1990).
\bibitem{Blum95} T. Blum and A. J. McKane,
  {\it Phys. Rev. E {\bf 52}}, 4741 (1995).
\bibitem{Bouchaud93} J-P. Bouchaud and M. E. Cates,
  {\it Phys. Rev. E {\bf 47}}, 1455 (1993).
  \\ (erratum), {Phys. Rev. E {\bf 48}}, 653 (1993).
\bibitem{Katzav02} E. Katzav and M. Schwartz,
 {\it Physica A {\bf 309}}, 69 (2002).
\bibitem{Colaiori01} F. Colaiori and M. A. Moore,
  {\it Phys. Rev. Lett. {\bf 86}}, 3946 (2001).
\bibitem{Cook90} J. Cook and B. Derrida,
 {\it J. Phys. A {\bf 23}}, 1523 (1990).
\bibitem{Marinari02} E. Marinari, A. Pagnani, G. Parisi and Z. R\'{a}cz,
  {\it Phys. Rev. E {\bf 65}}, 26136 (2002).
\bibitem{Tu94} Y. Tu,
  {\it Phys. Rev. Lett. {\bf 73}}, 3109 (1994).


\end{thebibliography}
\end{document}